\documentclass[prl,showpacs,preprintnumbers,amsmath,amssymb]{revtex4-1}
\usepackage{graphicx}
\usepackage{dcolumn}
\usepackage{amsmath}
\usepackage{bm}
\begin{document}
\title{\bf  {Hubble Diagrams in the Jordan and Einstein Frames}}
\author{Reza Rashidi \\
{\small Department of Physics, Shahid Rajaee Teacher Training University,
Tehran,  IRAN.}\\{\small E-mail: reza.rashidi@srttu.edu}}
 \begin{abstract}

Different models in cosmology generally predict different Hubble diagrams. Then, the comparison between the Hubble diagrams may be used as a way for distinguishing between different cosmological scenarios. But that is not always the case because there is no guarantee that two different models always have different Hubble diagrams. It may be possible for two physically-inequivalent models to have the same Hubble diagrams. In that case, the Hubble diagram cannot be used to differentiate between two models and it is necessary to find another way to distinguish between them. Therefore, the question of whether two different scenarios are distinguishable by using the Hubble diagrams is an important question which would not have an obvious answer. The Jordan and Einstein frames of $f(R)$ theories of gravity are inequivalent, provided that the metricity condition holds in both frames. In the present paper it is argued that if the time-variation of particle masses in the Einstein frame is taken into consideration, the Hubble diagram derived practically from type Ia supernova surveys does not enable us to differentiate between these two frames. Nevertheless, we show that by waiting long enough to measure the change in Hubble diagram it is possible to differentiate between two frames. In other words, the Hubble diagram cannot be employed alone to differentiate between two frames but comparison between the rates of changes in Hubble diagrams can provide a way to do so.
\end{abstract}
\maketitle
\section{Introduction}
The cosmological observations developed in the last two decades indicate that the universe is undergoing a phase of cosmic acceleration started after the matter domination \cite{1,2,3,4,5,6,7,8,9}. The simplest model which successfully explains the observational data is the $\Lambda$CDM ($\Lambda$-cold dark matter) model \cite{6,7,10}. This model also describes the energy budget of the universe and the formation of galaxies \cite{3,4,10}. However, the $\Lambda$CDM model is not problem-free \cite{11}. The main problem is that why the observed value of the cosmological constant is so small in comparison with the predicted vacuum energy of matter fields. Therefore, it seems that it is not possible to attribute it directly to the quantum vacuum energy and then the question arises as to what the origin of this cosmological constant is. There is also another problem called the cosmic coincidence problem which can be expressed as the question: why is the late-accelerating phase happening today that we are present to observe it? Because of these issues an alternative perspective towards resolving the acceleration problem has been proposed. The gravitational sector of Einstein's general relativity may not provide a sufficiently adequate description for gravitational interaction at very large scales. The motivation for modifying the gravitational part of the Einstein equation is not restricted to solving the cosmological problems. General relativity is not a renormalizable theory. Consequently, to quantize the gravitational fields conventionally, the Einstein-Hilbert action needs to be supplemented by higher order curvature terms \cite{12,13}. Also, in string theory and when quantum corrections are taken into account, the effective gravitational action at low energy level admits higher order curvature invariants \cite{14,15,16}. However, these quantum corrections to general relativity are often considered to be important at very small scales and therefore do not affect the gravitational phenomenology at large scales.

 There are many proposals to generalize the Einstein theory of gravity \cite{68,69,70} such as, for instance, the scalar-tensor theories \cite{17,18}, DGP (Dvali-Gabadadze-Porrati) gravity \cite{19}, brane-world gravity \cite{20}, TeVeS (Tensor-Vector-Scalar) theory \cite{21}, Einstein-Aether theory \cite{22} and $f(R)$ theories of gravity \cite{23,24,25,26,27,28,29}. Besides having the simplest form of the action, the $f(R)$ theories of gravity are ghost-free, provided that $f''(R)\geq 0$ \cite{23,30,31}, and can avoid the Ostrogradski instability \cite{32}.

 It is well-known that the action of an $f(R)$ theory in the presence of a matter field minimally coupled to the gravity can be transformed by a suitable conformal transformation to the Einstein-Hilbert action supplemented by a new dynamical scalar field which is minimally coupled to the gravity and non-minimally to the matter fields \cite{33,34,35,36,37,38,39,40}. The original set of variables is commonly called Jordan frame, while the transformed set, whose dynamics described by Einstein equations, is called Einstein frame \cite{39}. As a consequence of the non-minimal coupling between the new scalar field and matter field in the Einstein frame, the matter energy-momentum tensor is no longer covariantly conserved which can be attributed to a variation in masses. Since these two frames are conformally related they are mathematically equivalent (at least at the classical level), namely the space of solutions in one frame is isomorphic to another one. Here, an important question arises whether the physical contents of these frames are equivalent or not. On the other hand, each $f(R)$ theory of gravity is conformally related to a Brans-Dicke theory with the Brans-Dicke parameter $\omega_{0}=0$ \cite{40,41,42,23}. Therefore, if we take an arbitrary conformal transformation into account, the above question can be generalized to infinite number of conformally related frames \cite{43,44}. There are two different attitudes towards this issue. One of them asserts that the appearance of a new scalar field in the Einstein frame is only a field redefinition and a conformal transformation is nothing else a point dependent change in the units of measurement. For this reason, these two frames are dynamically equivalent \cite{65,66,43,44,45,46,47,48,74}. In other words, two dynamically equivalent theories are actually just different representations of the same theory. From this point of view, since the change of units in the Einstein frame implies that the length of a vector varies under parallel transport along a curve, one may assume that the metricity condition no longer holds and employs the Weyl-integrable geometry instead of the Riemannian geometry in this frame \cite{49,50,51}.

  In total contrast to the above interpretation, another approach is that the metricity condition can hold in both frames and then the units of measurement can be preserved constant in both of them.
 Thus, these two frames have essentially different physical contents (at least at a quantum level)\cite{29,39,52,53,54,55,56,57,58,59,60,64,71,72,73}.
  As mentioned in \cite{44}, in addition to the equations of motion, a set of interpretative rules assigned to the variables constitutes a fundamental part of a theory. In fact, the interpretative rules in a geometry with running units are different in comparison with the interpretative rules in a geometry with fixed units. Consequently, the two different Einstein frames, one equipped with the metricity condition and another one not, cannot be equivalent. In this context, another question arises that which one of these frames is the physical frame. There are many arguments in the literature to support the use of the Einstein frame versus its Jordan counterpart and vice versa \cite{39,57,58,59,60}. But, the aim of this paper is not to deal with this issue.

 In the present paper we shall focus on the following question. Assuming that the two frames are inequivalent (or equivalently assuming an Einstein frame with fixed units \cite{45}), is it practically possible to distinguish between them by analyzing the observational data obtained from type Ia supernova surveys? In other words, can the Hubble diagram of type Ia supernova reveal which one of the two frames correctly describes our universe? To investigate this problem we design and examine an example in detail and show that, by taking the time-variations of masses into account, the answer is not affirmative. In fact, it will be shown that the Hubble diagrams derived in practice from the observational data are exactly the same for two frames. Nevertheless, we will argue that it is not because of a physical equivalence between these frames and show that if we have enough time to observe the change in the Hubble diagram it is possible to distinguish between them. It means that the Hubble diagram cannot be employed alone to differentiate between two frames but comparison between the changes in Hubble diagrams can provide a way to distinguish between them. 

\section{Jordan and Einstein frames}
We assume that the space-time is four-dimensional and the signature of the metric is $(- + + +)$. Let us consider the action of a nonlinear (or $f(R)$) theory of gravity in the Jordan frame in which the matter fields (collectively denoted by $\psi$) are minimally coupled to gravity,
\begin{equation}\label{m1}
 S_{J}=\frac{1}{2}\int d^{4}x \sqrt{-g}f(R)+ S_{m}(g_{\mu\nu}, \psi),
\end{equation}
where $R$ and $g$ are respectively the Ricci scalar and the determinant of the metric and $f$ is a nonlinear function (we set $\hbar=c=8\pi G=1$). The Ricci scalar $R$ is defined by $R=g^{\mu\nu}R_{\mu\nu}$. The Ricci tensor $R_{\mu\nu}$ is
\begin{equation}\label{m2}
    R_{\mu\nu}=R^{\alpha}_{\mu\alpha\nu}= \partial_{\alpha}\Gamma^{\alpha}_{\mu\nu}-\partial_{\mu}\Gamma^{\alpha}_{\alpha\nu}
    +\Gamma^{\alpha}_{\mu\nu}\Gamma^{\rho}_{\rho\alpha}
    -\Gamma^{\alpha}_{\rho\nu}\Gamma^{\rho}_{\mu\alpha},
\end{equation}
where the connections $\Gamma^{\alpha}_{\mu\nu}$ are the metric connections defined as
\begin{equation}\label{m3}
  \Gamma^{\alpha}_{\mu\nu}=\frac{1}{2}g^{\alpha\rho}(\partial_{\mu}g_{\rho\nu}+\partial_{\nu}g_{\rho\mu}
  -\partial_{\rho}g_{\mu\nu}).
\end{equation}
It follows from the metricity condition
\begin{equation}\label{m4}
  \nabla_{\alpha}g_{\mu\nu}=\partial_{\alpha}g_{\mu\nu}-\Gamma^{\rho}_{\alpha\nu}g_{\rho\mu}
  -\Gamma^{\rho}_{\alpha\mu}g_{\rho\nu}=0.
\end{equation}
Varying the action (\ref{m1}) with respect to
the metric $g_{\mu\nu}$, one can derive the field equation
\begin{equation}\label{m5}
  f'(R)R_{\mu\nu}-\frac{1}{2} f(R)g_{\mu\nu}-(\nabla_{\mu}\nabla_{\nu}-g_{\mu\nu}\Box)f'(R)=T^{(m)}_{\mu\nu},
\end{equation}
where the prime denotes derivative with respect to the argument, $\nabla_{\mu}$ is the covariant derivative associated with the metric connection $\Gamma^{\alpha}_{\mu\nu}$, $\Box\equiv g^{\mu\nu}\nabla_{\mu}\nabla_{\nu}$ and $T^{(m)}_{\mu\nu}$ is the standard matter energy-momentum tensor
\begin{equation}\label{m6}
  T^{(m)}_{\mu\nu}=\frac{-2}{\sqrt{-g}}\frac{\delta S_{m}}{\delta g^{\mu\nu}}.
\end{equation}
The left hand side of equation (\ref{m5}) is divergence-free (generalized
Bianchi identity) implying the conservation equation
\begin{equation}\label{m7}
  \nabla^{\mu} T^{(m)}_{\mu\nu}=0.
\end{equation}
It may be viewed as a consequence of the invariance of the action (\ref{m1}) under diffeomorphisms.

Considering a dust fluid with $T^{(m)}_{\mu\nu}=\rho u_{\mu}u_{\nu}$ and $ g_{\mu\nu}u^{\mu}u^{\nu}=-1$, the conservation equation yields
\begin{equation}\label{m8}
 \nabla^{\mu}\rho u_{\mu}u_{\nu}+\rho\nabla^{\mu}u_{\mu}u_{\nu}+\rho u_{\mu}\nabla^{\mu}u_{\nu}=0.
\end{equation}
Multiplying both sides of this equation by $(u^{\nu}u^{\alpha}+g^{\nu\alpha})$ gives the geodesic equation
\begin{equation}\label{m9}
 u_{\mu}\nabla^{\mu}u_{\nu}=0.
\end{equation}
It means that a massive test particle follows geodesics of the metric in the Jordan frame and since the matter fields are minimally coupled to the gravity the Einstein equivalence principle is satisfied.

According to a well-known procedure \cite{23,33,34,35,36,37,38,39,40}, it is possible to represent the action (\ref{m1}) in the Einstein frame by means of a suitable conformal transformation. To derive it one can introduce a new dynamical field $\chi$ and the action
\begin{equation}\label{m10}
 S=\frac{1}{2}\int d^{4}x \sqrt{-g}[f(\chi)+f'(\chi)(R-\chi)]+ S_{m}(g_{\mu\nu}, \psi).
\end{equation}
Variation with respect to $\chi$ leads to the equation
\begin{equation}\label{m11}
  f''(\chi)(R-\chi)=0,
\end{equation}
which gives $R=\chi$, provided that $ f''(\chi)\neq 0$. Hence, the action (\ref{m10}) is dynamically equivalent to the action (\ref{m1}). Setting
\begin{equation}\label{m12}
  e^{2\phi/\sqrt{6}}\equiv f'(\chi) \hspace{.5cm},\hspace{.5cm} \tilde{g}_{\mu\nu}\equiv e^{2\phi/\sqrt{6}}g_{\mu\nu},
\end{equation}
and assuming that the change of variable $e^{2\phi/\sqrt{6}}= f'(\chi)$ is invertible, the action (\ref{m10}) takes the form
\begin{eqnarray}\label{m13}
 \nonumber S_{E}&=&\int d^{4}x\sqrt{-\tilde{g}}[\frac{\tilde{R}}{2}
  -\frac{1}{2}\tilde{g}^{\mu\nu}\partial_{\mu}\phi\partial_{\nu}\phi-U(\phi)] \\
   &+&S_{m}(e^{-2\phi/\sqrt{6}}\tilde{g}_{\mu\nu},\psi),
\end{eqnarray}
where the tilde means the quantities which are given in terms of the conformal metric $\tilde{g}_{\mu\nu}$ and
\begin{equation}\label{m14}
  U(\phi)=\frac{e^{2\phi/\sqrt{6}}\chi(\phi)-f(\chi(\phi))}{2e^{4\phi/\sqrt{6}}}.
\end{equation}
 The new fields $\tilde{g}_{\mu\nu}$ and $\phi$ provide the Einstein frame. The gravitational part of the action now contains only Einstein gravity. The field $\phi$ is minimally coupled to the gravity but non-minimally coupled to the matter field. This new field corresponds to the additional degree of freedom due to the higher order of the field equations in Jordan frame. Variation of the action (\ref{m13}) with respect to the new metric $\tilde{g}_{\mu\nu}$ leads to
 \begin{equation}\label{m15}
    \tilde{R}_{\mu\nu}-\frac{1}{2}\tilde{g}_{\mu\nu}\tilde{R}=\tilde{T}^{(\phi)}_{\mu\nu}+\tilde{T}^{(m)}_{\mu\nu},
 \end{equation}
 where
 \begin{equation}\label{m16}
  \tilde{T}^{(\phi)}_{\mu\nu}=\partial_{\mu}\phi\partial_{\nu}\phi
  -\frac{1}{2}\tilde{g}_{\mu\nu}\tilde{g}^{\alpha\beta}\partial_{\alpha}\phi\partial_{\beta}\phi
  -U(\phi)\tilde{g}_{\mu\nu},
 \end{equation}
 and
 \begin{equation}\label{m17}
  \tilde{T}^{(m)}_{\mu\nu}=\frac{-2}{\sqrt{-\tilde{g}}}\frac{\delta S_{m}}{\delta \tilde{g}^{\mu\nu}} \hspace{.2 cm}.
 \end{equation}
 Variation with respect to $\phi$ yields
 \begin{equation}\label{m18}
    \tilde{\Box}\phi-\frac{d U}{d\phi}=\frac{1}{\sqrt{6}}\tilde{g}^{\mu\nu}\tilde{T}^{(m)}_{\mu\nu}\equiv\frac{1}{\sqrt{6}}\tilde{T}^{(m)}
 \end{equation}
where $\tilde{\Box}\phi\equiv\tilde{g}^{\mu\nu}\tilde{\nabla}_{\mu}\tilde{\nabla}_{\nu}$ and $\tilde{\nabla}_{\mu}$ denotes the covariant derivative associated to the metric connection
\begin{equation}\label{m19}
 \tilde{\Gamma}^{\alpha}_{\mu\nu}=\frac{1}{2}\tilde{g}^{\alpha\rho}(\partial_{\mu}\tilde{g}_{\rho\nu}
 +\partial_{\nu}\tilde{g}_{\rho\mu}
  -\partial_{\rho}\tilde{g}_{\mu\nu}).
\end{equation}
 Taking the covariant derivative of equation (\ref{m15}) and using equation (\ref{m18}), one gets
 \begin{equation}\label{m20}
  \tilde{\nabla}^{\mu}\tilde{T}^{(m)}_{\mu\nu}=- \tilde{\nabla}^{\mu}\tilde{T}^{(\phi)}_{\mu\nu}= \frac{-1}{\sqrt{6}}\tilde{T}^{(m)}\partial_{\nu}\phi.
 \end{equation}
 The energy-momentum tensor of matter is no longer conserved due to the non-minimal coupling of the field $\phi$ to the matter field.
Setting $\tilde{T}^{(m)}_{\mu\nu}=\tilde{\rho} \tilde{u}_{\mu}\tilde{u}_{\nu}$ and $ \tilde{g}_{\mu\nu}\tilde{u}^{\mu}\tilde{u}^{\nu}=-1$, equation (\ref{m20}) implies
\begin{equation}\label{m21}
 \tilde{u}^{\mu}\tilde{\nabla}_{\mu}\tilde{u}_{\nu}=\frac{1}{\sqrt{6}}(\partial_{\nu}\phi
 +\partial_{\mu}\phi \tilde{u}^{\mu}\tilde{u}_{\nu}).
\end{equation}
It means that in the Einstein frame the world-lines of massive test particles do not satisfy the geodesic equation associated with the metric $\tilde{g}_{\mu\nu}$. The last term on the right hand side can be removed after a suitable reparametrization ($\tilde{u}_{\mu}\rightarrow e^{\phi/\sqrt{6}}\tilde{u}_{\mu}$), whereas the first term is not removable and describes the direct coupling of the field $\phi$ to matter fields in the Einstein frame.
A comparison between (\ref{m6}) and (\ref{m17}) shows that
\begin{equation}\label{m22}
 \tilde{T}^{(m)}_{\mu\nu}=e^{-2\phi/\sqrt{6}}T^{(m)}_{\mu\nu}.
\end{equation}
Hence, for a dust fluid we have
\begin{equation}\label{m23}
\tilde{\rho} \tilde{u}_{\mu}\tilde{u}_{\nu}=e^{-2\phi/\sqrt{6}}\rho u_{\mu}u_{\nu}.
\end{equation}
On the other hand, it is assumed that
\begin{equation}\label{m24}
\tilde{g}_{\mu\nu}\tilde{u}^{\mu}\tilde{u}^{\nu}= g_{\mu\nu}u^{\mu}u^{\nu}=-1,
\end{equation}
which implies
\begin{equation}\label{m25}
\frac{d x^{\mu}}{d\tilde{s}}=\tilde{u}^{\mu}=e^{-\phi/\sqrt{6}} u^{\mu}=e^{-\phi/\sqrt{6}}\frac{d x^{\mu}}{d s},
\end{equation}
where $s$ and $\tilde{s}$ are affine parameters in the Jordan and Einstein frames respectively.
Equation (\ref{m23}) in combination with the above relation yields
\begin{equation}\label{m26}
  \tilde{\rho}=e^{-4\phi/\sqrt{6}}\rho.
\end{equation}
Setting $\rho=\frac{dm}{dV}$ and $\tilde{\rho}=\frac{d\tilde{m}}{d\tilde{V}}$, one gets
\begin{equation}\label{m27}
 d\tilde{m}=e^{-\phi/\sqrt{6}}dm,
\end{equation}
and the equation of motion (\ref{m21}) can be then rewritten as
\begin{equation}\label{m28}
 \tilde{u}^{\mu}\tilde{\nabla}_{\mu}(\tilde{m}\tilde{u}_{\nu})=-\partial_{\nu}\tilde{m},
\end{equation}
where $\tilde{m}=e^{-\phi/\sqrt{6}}m$ with constant $m$.

It is needless to say that there are many classical theories of gravity in addition to $f(R)$ theories of gravity, including the Brans-Dicke theory, scalar tensor theories, classical Kaluza-Klein theories and so on, for which a conformal transformation maps the initial frame into the Einstein frame.
Obviously, at the classical level there is a mathematical equivalence between the original (Jordan) frame and the Einstein frame. The space of solutions of the theory in one frame is isomorphic to the space of solutions in the conformally related frame.
But, this mathematical equivalence implies nothing about their physical contents \cite{39,57,58,59,60}. There is a long-standing question about the relation between the physical contents of two conformally related frames. There are two different attitudes towards this issue. One attitude asserts that a conformal transformation is nothing else a point dependent change in the units of measurement and the physical contents of two conformally related frames are equivalent \cite{65}. One may consider the Jordan frame as a frame with constant units and the Einstein frame as a frame with running (or point-dependent) units or vice versa \cite{43,44,45,46,47,48}. To accommodate running units we can employ Weyl-integrable geometry instead of Riemannian geometry \cite{49,50,51}. In the Jordan frame as a frame with constant units it is assumed that the metricity condition (\ref{m4}) holds and it is then a Riemannian space.  But, in the Einstein frame with running units, since the length of a vector varies under parallel transport along a curve the metricity condition should be replaced by \cite{49,50,51}
\begin{equation}\label{m29}
 \hat{\nabla}_{\alpha}\tilde{g}_{\mu\nu}\equiv\partial_{\alpha}\tilde{g}_{\mu\nu}-\hat{\Gamma}^{\rho}_{\alpha\nu}\tilde{g}_{\rho\mu}
  -\hat{\Gamma}^{\rho}_{\alpha\mu}\tilde{g}_{\rho\nu}=\sqrt{\frac{2}{3}}\partial_{\alpha}\phi \tilde{g}_{\mu\nu},
\end{equation}
 which implies the following non-metric-compatible connection:
 \begin{equation}\label{m30}
\hat{\Gamma}^{\rho}_{\alpha\mu}=\tilde{\Gamma}^{\rho}_{\alpha\mu}-\frac{1}{\sqrt{6}}\{\delta^{\rho}_{\alpha}\partial_{\mu}\phi
+\delta^{\rho}_{\mu}\partial_{\alpha}\phi-\tilde{g}_{\alpha\mu}\tilde{g}^{\rho\beta}\partial_{\beta}\phi\},
 \end{equation}
 where $\tilde{\Gamma}^{\rho}_{\alpha\mu}$ is the metric-compatible connection (\ref{m19}). The Einstein frame with the connection (\ref{m30}) is a Weyl-integrable space. It is not difficult to see that the Weyl connection of the Einstein frame is equal to the Riemannian connection of the Jordan frame, i.e. $\hat{\Gamma}^{\rho}_{\alpha\mu}=\Gamma^{\rho}_{\alpha\mu}$. It means that the rule of parallel transport is actually defined by the Jordan metric $g_{\mu\nu}$ and the Einstein metric $\tilde{g}_{\mu\nu}$ only determines the change in the units of measurement. Using equations (\ref{m9}), (\ref{m25}) and (\ref{m27}), one obtains
 \begin{equation}\label{m31}
   \tilde{u}^{\mu}\hat{\nabla}_{\mu}(\tilde{m}\tilde{u}_{\nu})=0.
 \end{equation}
It means that the origin of deviation from geodesic equation in the Einstein frame with Weyl-integrable geometry is the point-dependent units of mass $\tilde{m}$ and proper time $d\tilde{s}$.

In contrast to the above approach, another interpretation is that the measurement units in both Jordan frame and Einstein frame are constant and the metricity condition (\ref{m4}) holds in both frames \cite{49,50,57,58,59,60}, hence both frames are Riemannian spaces. In this context, the deviation from geodesic equation is caused by the fifth force generated by the field $\phi$. Equation (\ref{m28}) suggests that a massive particle deviates from geodesics of the Einstein frame because its mass is a function of the space-time points. However, this equation of motion does not violate the weak equivalence principle because it is assumed that the values of the field $\phi$ for all types of particles are the same in each point of space-time, and accordingly the trajectories of test particles are independent of their masses. But, this is not the case for extended objects (Dicke-Nordtvedt effect). On the other hand, the equation of motion (\ref{m21}) implies that the acceleration of a freely-falling test particle in a fixed point with respect to a local inertial observer moving on a geodesic of the Einstein frame depends on their relative velocity. Consequently, for instance, hot bodies have different accelerations in comparison with cold identical ones \cite{61} or atoms with different spin states have different accelerations \cite{62}. Thus, in the Einstein frame, the Universality of Free Fall is violated for inertial observers. A similar phenomenon also occurs in General Relativity (GR) and in the Jordan frame but not for inertial observers. For example, in the Schwarzschild space-time the acceleration of a freely-falling test particle with respect to a local static observer who is at rest relative to the gravitational source depends on their relative velocity. Obviously, in GR and also in the Jordan frame for freely-falling observers which follow geodesics the Universality of Free Fall and local Lorentz invariance are satisfied.

  It is worthwhile to note that there is another important difference between the Jordan and Einstein frames. Actually, in the Jordan frame the freely-falling observers (who not acted upon by non-gravitational forces) and the inertial observers (who follow geodesics) are equivalent. But it is not the case for the Einstein frame because the field $\phi$ is a universal field and coupled to all types of matter including the measuring instruments, and therefore to force an object into following a geodesic it is necessary to exert a non-gravitational force on it to eliminate the force generated by the field $\phi$. In other words, an inertial observer must be generally affected by a non-gravitational force. It means that the natural (or normal) motion is no longer the following of geodesics. Thus, one may distinguish between inertial and freely-falling observers in the Einstein frame and defines an inertial observer as an observer who moves on a geodesics and a freely-falling observer as an observer who is always at rest relative to a freely-falling test particle (the one not acted on by a well-known non-gravitational forces such as electromagnetic force).

  So the question arises whether or not the Universality of Free Fall is violated for freely-falling observers.
  The following argument shows that the Universality of Free Fall is also violated for freely-falling observers in the Einstein frame.
  Equation (\ref{m28}) can be expressed as
  \begin{equation}\label{a1}
     \tilde{u}^{\mu}\tilde{\nabla}_{\mu}(\tilde{u}_{\nu})=
     -\frac{\partial_{\nu}\tilde{m}}{\tilde{m}}-\frac{\partial_{\mu}\tilde{m}}{\tilde{m}}\tilde{u}^{\mu}\tilde{u}_{\nu}\equiv a_{\nu}.
  \end{equation}
  Contracting the left hand side with $\tilde{u}^{\nu}$ identically yields zero, i.e. $\tilde{u}^{\nu}a_{\nu}=0$. It is then possible to employ the Fermi-Walker (or Fermi normal) coordinates near each curve that satisfies the above equation \cite{67}. We use $x^{\mu}$ to denote the Fermi-Walker coordinates near the world-line of a freely-falling observer. Then the world-line of the observer can be described by $d x^{0}=dt$ and $x^{i}=0$, where $dt$ is the proper time along the world-line and $i=1,2,3$. In these coordinates, the metric near the world-line can be expressed as \cite{67}
  \begin{eqnarray}\label{a2}
    \nonumber \tilde{g}_{00}&=&-(1+2a_{i}(t)x^{i}+(a_{i}(t)x^{i})^2+\tilde{R}_{0i0j}(t)x^{i}x^{j}+\mathcal{O}(3)),\\
    \nonumber \tilde{g}_{0i}&=&-\frac{2}{3}\tilde{R}_{0jik}(t)x^{j}x^{k}+\mathcal{O}(3),\\
    \tilde{g}_{ij}&=&\delta_{ij}-\frac{1}{3}\tilde{R}_{ikjm}(t)x^{k}x^{m}+\mathcal{O}(3),
  \end{eqnarray}
  where $a_{i}(t)$ are the spatial components of $a_{\nu}$ on the world-line and $\tilde{R}_{\mu\nu\alpha\beta}(t)$ are the components of the Riemann tensor evaluated again on the world-line. It is no difficult to see that the metric reduces to $\tilde{g}_{00}=-1$ and $\tilde{g}_{ij}=\delta_{ij}$ on the world-line and the non-vanishing Christoffel symbols on the world-line are $\tilde{\Gamma}^{0}_{0i}=\tilde{\Gamma}^{i}_{00}=a_{i}(t)$. Since the four-velocity $\tilde{u}^{\mu}$ on the world-line is $(1,0,0,0)$ we get
  \begin{equation}\label{a3}
    a_{i}(t)=-\frac{\partial_{i}\tilde{m}(x^0=t,x^i=0)}{\tilde{m}(x^0=t,x^i=0)}.
  \end{equation}
 Now consider another freely-falling test particle whose velocity relative to our observer does not vanish when its world-line intersects the world-line of the observer. Near the observer the world-line of this particle can be described by the Fermi-Walker coordinates as $x^{\mu}(s)$, where $s$ is the proper time along the particle world-line. The acceleration of the particle with respect to the observer at the intersection point is
 \begin{eqnarray}\label{a4}
   A^{i}&\equiv & \frac{d^2 x^i(s)}{(d x^0(s))^2}=\frac{d^2 x^i(s)}{d t^2}\\
  \nonumber  &=&-(\frac{d s}{d t})^2\frac{d^2t}{d s^2}\frac{d x^i}{d t}+(\frac{d s}{d t})^2\frac{d^2 x^i}{d s^2}.
 \end{eqnarray}
Assuming that the velocity of the particle relative to the observer is $\frac{d x^i}{d t}=v^i$ at the intersection point, we have
\begin{equation}\label{a5}
 (\frac{d s}{d t})^2=(\frac{d s}{d x^0})^2=1-v^2,
\end{equation}
where $v^2=v^iv_{i}$. But, the world-line of the particle must also satisfy the equation of motion (\ref{a1}). Since the intersection point is also on the observer world-line and on this world-line the only non-vanishing Christoffel symbols are $\tilde{\Gamma}^{0}_{0i}=\tilde{\Gamma}^{i}_{00}=a_{i}(t)$, equation (\ref{a1}) yields
\begin{eqnarray}\label{a6}
  \nonumber  \frac{d^2 t}{d s^2}=\frac{d^2 x^0}{d s^2}=&-&2a_{j}\frac{d x^0}{d s}\frac{d x^j}{d s}-\frac{d x^0}{d s}\frac{d x^{\mu}}{d s}\frac{\partial_{\mu} \tilde{m}}{\tilde{m}}\\
    &-&\frac{\partial^0 \tilde{m}}{\tilde{m}},
\end{eqnarray}
and
\begin{equation}\label{a7}
    \frac{d^2 x^i}{d s^2}=-a^i (\frac{d x^0}{d s})^2-\frac{d x^i}{d s}\frac{d x^{\mu}}{d s}\frac{\partial_{\mu} \tilde{m}}{\tilde{m}}-\frac{\partial^i \tilde{m}}{\tilde{m}}.
\end{equation}
Taking equations (\ref{a3}) and (\ref{a5}) into account, one can obtain
\begin{equation}\label{a8}
    \frac{d^2 t}{d s^2}=-\frac{v^2}{1-v^2}\frac{\partial_{0} \tilde{m}}{\tilde{m}}+\frac{v^j}{1-v^2}\frac{\partial_{j} \tilde{m}}{\tilde{m}} ,
\end{equation}
and
\begin{equation}\label{a9}
    \frac{d^2 x^i}{d s^2}=-\frac{v^i}{1-v^2}\frac{\partial_{0} \tilde{m}}{\tilde{m}}+(\frac{v^2}{1-v^2}\delta^{ij}-\frac{v^iv^j}{1-v^2})\frac{\partial_{j} \tilde{m}}{\tilde{m}} .
\end{equation}
Substituting the relations (\ref{a5}), (\ref{a8}) and (\ref{a9}) in equation (\ref{a4}), we find
 \begin{equation}\label{a10}
  A^i=-(1-v^2)v^i\frac{\partial_{0} \tilde{m}}{\tilde{m}}+(v^2\delta^{ij}-2v^i v^j)\frac{\partial_{j} \tilde{m}}{\tilde{m}}.
 \end{equation}
 It is not difficult to see that not only the acceleration of particle with respect to the observer does not vanish, but also freely-falling test particles, with different velocities relative to the freely-falling observer, have different accelerations with respect to the observer.
It means that the Universality of Free Fall is violated for freely-falling observers in the Einstein frame; whereas, in the Jordan frame in contrast to the Einstein frame, the acceleration of all freely-falling particles is always zero with respect to a freely-falling observer.

\section{Hubble Diagrams}
As mentioned in the previous section, physics laws (the relations between physical observables) for freely-falling observers in the Jordan frame generally differ from the ones in the Einstein frame, provided that the metricity condition holds in both frames.
This raises the question as to whether or not the cosmological red-shift surveys (i.e. the recent observational data obtained from type Ia supernova surveys) can practically reveal which one of the two frames correctly describes our universe. In other words, is it possible to distinguish between the two frames by analyzing these observational data?

The key observational quantities in cosmology are the brightness (i.e. energy flux density of an astronomical object) and the received light wavelength. Using these observational data, the Hubble diagram (the plot of proper (or luminosity) distance versus the redshift) can be derived. The Hubble curve can be used to distinguish between different cosmological scenarios. To derive this diagram, one should compare the received wavelength with the emitted wavelength to determine the wavelength shift. But, it is not possible for us to examine the emitted light directly; therefore, we have to compare the received wavelength with the wavelength measured in a laboratory on Earth. Additionally, to determine the proper distance of a very distant astronomical object it is required to know its intrinsic luminosity (i.e. emitted power) because direct methods such as the parallax method are not applicable. But again, we do not have any access to the light source and cannot measure its intrinsic luminosity directly. One of the conventional approaches to determine large distances is based on using the standard candles. A standard candle is an astronomical object that has a known intrinsic luminosity. To calibrate a standard candle one has to measure its brightness and its proper distance by means of the parallax method and then the intrinsic luminosity can be computed using the inverse square law. Consequently, this calibration works only at close distances wherein the parallax method is applicable and the redshift parameter is negligible. In other words, we can compare the features of a very distant source, such as its brightness and light wavelength, only with the features of the same objects in our vicinity. The use of this method for deriving the Hubble diagram can be justified if physics laws in our vicinity are the same as ones in a long distance. Therefore, this method can be employed to investigate the $f(R)$ theories of gravity and also the theory of general relativity. But for another cases such as theories wherein particle masses change; although, such a diagram can be plotted, we have to care about its interpretation.

Although, the process of deriving the Hubble diagram in the Jordan frame is well known and resembles the one in general relativity we work it in detail because we want to compare it with its counterpart in the Einstein frame.
The line element of a homogeneous and isotropic universe in the Jordan frame of $f(R)$ theories of gravity can be expressed in terms of the comoving coordinates as
\begin{equation}\label{m32}
  ds^2=-dt^2+a^2(t)(\frac{dr^2}{1-kr^2}+r^2d\theta^2+r^2 \sin^2\theta d\varphi^2),
\end{equation}
where $a(t)$ is the scale factor and $k$ is the spatial curvature parameter. Taking the above line element into account, the world-lines defined by $dr=0,d\theta=d\varphi=0$ and $dt=ds$, where $ds$ is the proper time, satisfy the geodesic equation (\ref{m9}). So, they are the trajectories of inertial observers being at rest with respect to the coordinate system. Now, consider a light source, for simplicity, including $\Delta N$ excited Hydrogen atoms being at rest with respect to the coordinate of system and located at the coordinates $r=r_{em}$, $\theta=\pi/2$ and $\phi=0$. Assume that all atoms collapse to the ground state and the source emits an isotropic electromagnetic signal consisting of $\Delta N$ monochrome photons during the time interval $[t_{em}, t_{em}+\Delta t_{em}]$. Consequently, its intrinsic luminosity measured by a comoving local observer is
\begin{equation}\label{m33}
  \mathfrak{L}_{em}= \frac{h c\Delta N}{\lambda_{em}\Delta t_{em}},
\end{equation}
where $h$ and $c$ are respectively the Planck constant and velocity of light and $\lambda_{em}$ is the wavelength of emitted photons measured by the comoving local observer. Let us consider another inertial observer being at rest with respect to the coordinate system and located at the coordinates $r=r_{rec}=0$, $\theta=\pi/2$ and $\phi=0$. Assuming that this observer receives the mentioned signal during the time interval $[t_{rec}, t_{rec}+\Delta t_{rec}]$, we have
\begin{equation}\label{m35}
  \Delta t_{rec}=\frac{a_{rec}}{a_{em}}\Delta t_{em}
\end{equation}
and the wavelength of received photons is
\begin{equation}\label{m34}
 \lambda_{rec}=\frac{a_{rec}}{a_{em}}\lambda_{em},
\end{equation}
where $a_{em}$ and $a_{rec}$ are the scale factor at the emission and received time, respectively. To determine the wavelength shift the second observer needs to measure the wavelength $\lambda_{em}$. But the light source is not accessible to this observer and then she has to go to her laboratory and measure the wavelength of a photon created by a Hydrogen atom. Since the physics laws and the particle masses do not change in the Jordan frame, she can deduce that the wavelength $\lambda_{lab}$ which is measured in the laboratory is equal to $\lambda_{em}$. It gives
\begin{equation}\label{m36}
  1+z=\frac{\lambda_{rec}}{\lambda_{em}}=\frac{\Delta t_{rec}}{\Delta t_{em}}=\frac{\lambda_{rec}}{\lambda_{lab}},
\end{equation}
where $z$ is the wavelength shift. Another quantity which can be measured by the second observer is the brightness (energy flux) of the source. It can be expressed as
\begin{eqnarray}\label{m37}
  f_{rec}&=&\frac{h c\Delta N}{A(D_{p})\lambda_{rec}\Delta t_{rec}}\\
  \nonumber &=&\frac{h c\Delta N}{4\pi(\frac{a_{rec}}{\sqrt{k}}\sin(\frac{\sqrt{k}}{a_{rec}}D_{p}))^{2} \lambda_{rec}\Delta t_{rec}},
\end{eqnarray}
where $D_{p}$ is the proper (physical) distance between the observer and the source at the received time $t_{rec}$ and $A(D_{p})$ is the surface area of the sphere with center at $r_{em}$ and radius $D_{p}$. Combining equations (\ref{m33}) and (\ref{m36}) with the above equation yields
\begin{equation}\label{m38}
 (\frac{a_{rec}}{\sqrt{k}}\sin(\frac{\sqrt{k}}{a_{rec}}D_{p}))^{2}=\frac{\mathfrak{L}_{em}}{4\pi f_{rec}(1+z)^2}.
\end{equation}
Therefore, to calculate the proper distance it is required to know the intrinsic luminosity of the source or generally we have to know the internal structure of the source which for the present example is the number of Hydrogen atoms $\Delta N$. Here, we need to use a standard candle. In practice, it means that we have to find in our vicinity an object resembling the light source. Measuring the brightness of this local source and its distance by means of the parallax method, its intrinsic luminosity can be determined and it is justifiable to set $\mathfrak{L}_{em}=\mathfrak{L}_{loc}$. One can then apply this method to various light sources in different distances and plot the proper distance $D_{p}$ versus the wavelength shift $z$ (the Hubble diagram).

On the other hand, using the line element (\ref{m32}) for a light ray connecting the source and second observer, it follows that \cite{63}
\begin{equation}\label{m39}
  D_{p}(z,t_{rec})=- a_{rec}\int_{t_{rec}}^{t_{em}}\frac{dt}{a(t)}=\int^{z}_{0}\frac{dz'}{H(z',t_{rec})},
\end{equation}
where $H$ is the Hubble parameter, i.e. $H=\frac{da/dt}{a}$ and $1+z'=\frac{a_{rec}}{a(t)}$. This is the key relation that, by means of it, the theoretical predictions can be compared with the observational data. The left hand side is the Hubble diagram and the right one comes from our theoretical model. Thus, deriving the Hubble diagram and its comparison with the model in the Jordan frame is a straightforward process.

But, the situation in the Einstein frame is completely different. In the corresponding Einstein frame the line element (\ref{m32}) becomes
\begin{eqnarray}\label{m40}
 d\tilde{s}^2&=&\Omega^2(t)ds^2\\
 \nonumber &=&-d\tilde{t}^2+\tilde{a}^2(t)(\frac{dr^2}{1-kr^2}+r^2d\theta^2+r^2 \sin^2\theta d\varphi^2),
\end{eqnarray}
where $d\tilde{t}=\Omega(t)dt$, $\tilde{a}=\Omega(t)a(t)$ and $\Omega(t)$ is the conformal factor, i.e. $\Omega(t)=e^{(\phi/\sqrt{6})}$. The world-lines defined by $dr=0,d\theta=d\varphi=0$ and $d\tilde{t}=d\tilde{s}$, where $d\tilde{s}$ is the proper time in the Einstein frame, satisfy equation (\ref{m21}). Since the right hand side of equation (\ref{m21}) always vanishes on these world-lines, it is then possible to regard these world-lines as the trajectories of not only freely-falling observers but also inertial observers. In other words, the freely-falling observers and the inertial observers coincide on these world-lines. Let us now study the above mentioned example in the Einstein frame. The intrinsic luminosity of the source is then
\begin{equation}\label{m41}
  \mathfrak{\tilde{L}}_{em}= \frac{h c\Delta N}{\tilde{\lambda}_{em}\Delta\tilde{t}_{em}},
\end{equation}
where $\tilde{\lambda}_{em}$ is the wavelength of the light source in the Einstein frame and $\Delta\tilde{t}_{em}$ is the length of the proper time interval of emission. The second observer receives the light with the wavelength
\begin{equation}\label{m42}
  \tilde{\lambda}_{rec}=\frac{\Omega_{rec}a_{rec}}{\Omega_{em}a_{em}}\tilde{\lambda}_{em},
\end{equation}
where $\Omega_{em}$ and $\Omega_{rec}$ are respectively the conformal factors at the emission and received time. The change in time interval is
\begin{equation}\label{m43}
  \frac{\Delta\tilde{t}_{rec}}{\Delta\tilde{t}_{em}}=\frac{\Omega_{rec}a_{rec}}{\Omega_{em}a_{em}},
\end{equation}
where $\Delta\tilde{t}_{rec}$ is the received proper time interval. To determine the wavelength shift, the second observer has to measure the wavelength created by a Hydrogen atom in the laboratory. But in the Einstein frame the wavelength in the laboratory is no longer equal to the emitted wavelength because the particle masses are not constant. According to the fact that the wavelength of the light given off by a hydrogen atom is proportional to the inverse of the hydrogen reduced mass, equation (\ref{m27}) yields
\begin{equation}\label{m44}
  \frac{\tilde{\lambda}_{lab}}{\tilde{\lambda}_{em}}=\frac{\Omega_{rec}}{\Omega_{em}},
\end{equation}
where $\tilde{\lambda}_{lab}$ is the wavelength measured by the second observer in the lab. Since it is impossible for the second observer to practically measure the wavelength $\tilde{\lambda}_{em}$, the quantity that can be determined in practice is
\begin{equation}\label{m45}
 \frac{\tilde{\lambda}_{rec}}{\tilde{\lambda}_{lab}}=\frac{\Omega_{em}}{\Omega_{rec}}(1+\tilde{z})=\frac{a_{rec}}{a_{em}}=1+z,
\end{equation}
where $\tilde{z}=\frac{\tilde{\lambda}_{rec}}{\tilde{\lambda}_{em}}-1$. It means that the quantity which is actually measurable in the Einstein frame is the wavelength shift of the Jordan frame, i.e. $z$, and not $\tilde{z}$. The brightness measured by the second observer in the Einstein frame is
\begin{eqnarray}\label{m46}
\tilde{f}_{rec}&=&\frac{h c\Delta N}{\tilde{A}(\tilde{D}_{p})\tilde{\lambda}_{rec}\Delta \tilde{t}_{rec}}\\
\nonumber &=& \frac{\tilde{\mathfrak{L}}_{em}}{4\pi (\frac{\tilde{a}_{rec}}{\sqrt{k}}\sin(\frac{\sqrt{k}}{\tilde{a}_{rec}}\tilde{D}_{p}))^{2}(1+\tilde{z})^2},
\end{eqnarray}
where $\tilde{D}_{p}$ is the proper distance between the light source and the second observer at the received time. The expression following the second equality is obtained by using equations (\ref{m41}), (\ref{m43}) and (\ref{m45}). Hence, according to equation (\ref{m45}), we have
\begin{equation}\label{m47}
(\frac{\tilde{a}_{rec}}{\sqrt{k}}\sin(\frac{\sqrt{k}}{\tilde{a}_{rec}}\tilde{D}_{p}))^{2}=(\frac{\Omega_{em}}{\Omega_{rec}})^2 \frac{\tilde{\mathfrak{L}}_{em}}{4\pi \tilde{f}_{rec}(1+z)^2}.
\end{equation}
But, in contrast to the Jordan frame, the luminosity $\tilde{\mathfrak{L}}_{em}$ is no longer equal to the luminosity of a local standard candle in the Einstein frame. The luminosity of a local source can be expressed as
\begin{equation}\label{m48}
\mathfrak{\tilde{L}}_{loc}= \frac{h c\Delta N}{\tilde{\lambda}_{loc}\Delta \tilde{t}_{loc}}.
\end{equation}
Since particle masses are no longer constant in the Einstein frame the wavelength $\tilde{\lambda}_{loc}$ and the local emission time $\Delta \tilde{t}_{loc}$ differ respectively from the emission wavelength $\tilde{\lambda}_{em}$ and the emission time $\Delta \tilde{t}_{em}$. The emission times are different because the transition rate between energy eigenstates is proportional to the mass (the Fermi golden rule). Thus, it follows from equation (\ref{m27}) that
\begin{equation}\label{m49}
  \frac{\Delta \tilde{t}_{em}}{\Delta \tilde{t}_{loc}}=\frac{\tilde{\lambda}_{em}}{\tilde{\lambda}_{loc}}=\frac{\Omega_{em}}{\Omega_{rec}}.
\end{equation}
Substituting equation (\ref{m49}) in equation (\ref{m48}) and comparing it with equation (\ref{m41}), we find that
\begin{equation}\label{m50}
  \frac{\mathfrak{\tilde{L}}_{em}}{\mathfrak{\tilde{L}}_{loc}}=(\frac{\Omega_{rec}}{\Omega_{em}})^2.
\end{equation}
Then by equation (\ref{m47}) we have
\begin{equation}\label{m51}
(\frac{\tilde{a}_{rec}}{\sqrt{k}}\sin(\frac{\sqrt{k}}{\tilde{a}_{rec}}\tilde{D}_{p}))^{2}=\frac{\tilde{\mathfrak{L}}_{loc}}{4\pi \tilde{f}_{rec}(1+z)^2}.
\end{equation}
Since all quantities on the right hand side can be measured by the second observer, it is possible in practice to plot a graph of $\tilde{D}_{p}$ against $z$. Now the question arises whether or not the comparison of this graph and the theoretical predictions can make a distinction between the Jordan and Einstein frame. Obviously, the graph of $\tilde{D}_{p}$ against $\tilde{z}$ (the Hubble diagram in Einstein frame) differs from the graph of $D_{p}$ against $z$ (the Hubble diagram in Jordan frame) and one can then distinguish between these two frames. But, in practice, we can not construct the Hubble diagram in the Einstein frame and can only derive the graph of $\tilde{D}_{p}$ versus $z$ from attainable data. Comparing the line element (\ref{m32}) and the line element (\ref{m40}), we get
\begin{equation}\label{m52}
 \tilde{D}_{p}=\Omega_{rec} D_{p}.
\end{equation}
 Then from equation (\ref{m39}), it follows that
 \begin{equation}\label{m53}
 \tilde{D}_{p}=\Omega_{rec}\int^{z}_{0}\frac{dz'}{H(z')}.
 \end{equation}
Then, according to equations (\ref{m38}), (\ref{m39}) and (\ref{m51}), one finds that
\begin{equation}\label{m54}
 (\frac{\tilde{\mathfrak{L}}_{loc}}{4\pi \tilde{f}_{rec}(1+z)^2})^{\frac{1}{2}}=\frac{\tilde{a}_{rec}}{\sqrt{k}}\sin(\frac{\sqrt{k}}{a_{rec}}\int^{z}_{0}\frac{dz'}{H(z')}),
  \end{equation}
  and
  \begin{equation}\label{m55}
  (\frac{\mathfrak{L}_{loc}}{4\pi f_{rec}(1+z)^2})^{\frac{1}{2}}= \frac{a_{rec}}{\sqrt{k}}\sin(\frac{\sqrt{k}}{a_{rec}}\int^{z}_{0}\frac{dz'}{H(z')}).
 \end{equation}
The values of the left hand sides of these equations are determined by observational data while the values of the right hand sides come from our models. If the values of the left and right hand sides in the first equation coincide, the Einstein frame becomes an acceptable model and if it occurs for the second equation, the Jordan frame is acceptable. But, we can always set $\Omega_{rec}=1$ by applying a constant conformal transformation. This is possible because a constant conformal transformation can be regarded as a coordinate transformation and the Einstein frame is invariant under coordinate transformations. If we set $\Omega_{rec}=1$, the right hand sides of both equations become equal. It means that there is not practically any difference between the Hubble diagrams extracted from observational data in both frames. Therefore, it is impossible to distinguish between two frames by comparison of  equations (\ref{m54}) and (\ref{m55}). The above result can be also generalized to all conformally related frames. If the observational data confirms the theoretical predictions, not only the Jordan frame is acceptable but also all conformally related frames including the Einstein frame are acceptable models. Conversely, if there appears to be a discrepancy between theory and observation, all conformally related frames fail.

This result does not arise from the physical equivalence of the frames because we assume that the metricity condition holds in two frames and therefore, as was mentioned in the previous section, these frames can not be generally equivalent. In fact, if we have enough time to observe the change in the Hubble diagram, it may be possible to distinguish between these frames. The proper distance $D_{p}$, according to equation (\ref{m39}), can be express as
\begin{equation}\label{m56}
 D_{p}=-a(t_{rec})\int_{t_{rec}}^{t_{em}}\frac{dt}{a(t)}.
 \end{equation}
It shows that the proper distance is actually a function of two parameters $t_{rec}$ and $t_{em}$. Taking the relation $1+z=\frac{a(t_{rec})}{a(t_{em})}$ into account, one may replace the variable $t_{em}$ by $z$, then the proper distance becomes a function of $t_{rec}$ and $z$. In other words, the Hubble diagram can change with respect to the time $t_{rec}$. Holding $z$ constant, the partial derivative of $D_{p}$ with respect to $t_{rec}$ represents the rate of change of the Hubble diagram in the Jordan frame for the fixed $z$. It should be noted that holding $z$ constant does not mean that a particular supernova has a constant redshift. To hold $z$ constant at two different times one should measure the redshifts of two different supernovae. In other words, when $z$ is held fixed and $t_{rec}$ changes, $D_{p}$ as a function of $z$ and $t_{rec}$ does not represent the change in the proper distance of a particular supernova. In fact, it represents the proper distances of supernovae which have the same redshifts at different times.

 On the other hand, according to equation (\ref{m52}) and setting $\Omega(t_{rec})=1$, the rate of change of the Hubble diagram in the Einstein frame is
\begin{eqnarray}\label{m57}
 \frac{\partial}{\partial\tilde{t}_{rec}}\tilde{D}_{p}(z,t_{rec})&=&\frac{\partial}{\partial t_{rec}}\tilde{D}_{p}(z,t_{rec})\\
 &=&\frac{d \Omega(t_{rec})}{d t_{rec}}D_{p}(z,t_{rec})+\frac{\partial}{\partial t_{rec}}D_{p}(z,t_{rec})\nonumber,
\end{eqnarray}
which differs from the rate of change in the Jordan frame. The comparison between these changes can be employed to distinguish between two frames. It should be noted that this difference is not a result of the time rescaling at the received time because we have set $\Omega(t_{rec})=1$ and then the unites of time in two frames are the same ($d \tilde{t}_{rec}=\Omega(t_{rec})dt_{rec}=dt_{rec} $).
As an example, let us assume that $a=H_{0}t$ and the spatial curvature $k$ vanishes; then the proper distance $D_{p}$ can be express as
\begin{equation}\label{m58}
 D_{p}(z, t_{rec})=-a(t_{rec})\int_{t_{rec}}^{t_{em}}\frac{dt}{a(t)}=t_{rec}\ln(1+z).
 \end{equation}
 The partial derivative of $D_{p}$ with respect to $t_{rec}$ is independent of $t_{rec}$. It means that the rate of change of the Hubble diagram in the Jordan frame is independent of time. On the other hand, since $\frac{d R}{d t}\neq 0$ ($R$ denotes the Ricci scalar) we have $\frac{d\Omega(t_{rec})}{d t_{rec}}\neq 0$ and hence $\frac{\partial\tilde{D}_{p}}{\partial \tilde{t}_{rec}}$, according to equation (\ref{m57}), can be dependent on time. In other words, the rate of change of the Hubble diagram in the Einstein frame depends on time. Thus, it is possible to differentiate between the two frames.

It is worthwhile to note that the result of the present paper is in conflict with the one derived in \cite{59}. In that paper, it is concluded that the Hubble diagrams of two frames are different. This results from the assumption that the redshift parameter $\tilde{z}$ is measurable; whereas, we argued that if one takes the variation of masses into account, this parameter is not practically measurable. Here, one may criticize that the variation in masses is not an unavoidable feature of the Einstein frame. For example, one may attribute the right hand side of equation (\ref{m21}) to the fifth force without considering any change in masses. But, we show that it may not be true. Taking equations (\ref{m20}), (\ref{m21}) and the line element (\ref{m40}) into account and setting $\tilde{T}^{(m)}_{\mu\nu}=\tilde{\rho} \tilde{u}_{\mu}\tilde{u}_{\nu}$ and $\tilde{u}_{\mu}=(1,0,0,0)$, it follows that
\begin{equation}\label{m59}
  \frac{\partial\tilde{\rho}}{\partial\tilde{t}}+ 3\frac{\tilde{\rho}}{\tilde{a}}\frac{d\tilde{a}}{d\tilde{t}}=-\frac{1}{\sqrt{6}}\frac{d\phi}{d\tilde{t}}\rho.
\end{equation}
The mass density $\tilde{\rho}$ can be expressed as $\tilde{\rho}\sim\frac{\tilde{m} N}{\tilde{a}^3}$, where $N$ is the total number of particles which is assumed to be fixed and $\tilde{m}$ is the rest mass of one particle. Hence we have
\begin{equation}\label{m60}
 \frac{\partial\tilde{m}}{\partial\tilde{t}}=-\frac{1}{\sqrt{6}}\frac{d\phi}{d\tilde{t}},
\end{equation}
which is compatible with equation (\ref{m27}) and indicates that the variation in masses is an inevitable feature of the Einstein frame.


\begin{thebibliography}{99}

\bibitem{1}A. G. Riess, et al. High-z Supernova Search Team, Observational evidence from supernovae for an accelerating universe and a cosmological constant, Astron. J. 116(1998)1009.
\bibitem{2}S. Perlmutter, et al. Supernova Cosmology Project, Measurements of omega and lambda from 42 high redshift supernovae, Astrophys. J. 517(1999)565.
\bibitem{3}D. J. Eisenstein, et al. (SDSS), Detection of the Baryon Acoustic Peak in the Large-Scale Correlation Function of SDSS Luminous Red Galaxies, Astrophys. J. 633(2005)560.
\bibitem{4}P. Astier, et al. (The SNLS), The Supernova Legacy Survey: measurement of  $\Omega_{M}$, $\Omega_{\Lambda}$ and $w$ from the first year data set, Astron. Astrophys. 447(2006)31.
\bibitem{5}D. N. Spergel, et al. (WMAP), Three-Year Wilkinson Microwave Anisotropy Probe (WMAP) Observations: Implications for Cosmology, Astrophys. J. Suppl. 170(2007)377.
\bibitem{6}L. Perivolaropoulos, Accelerating Universe: Observational Status and Theoretical Implications, arXiv:astro-ph/0601014.
\bibitem{7}H. Jassal, J. Bagla and T. Padmanabhan, Observational constraints on low redshift evolution of dark energy: How consistent are different observations?, Phys.Rev.D 72(2005)103503 [arXiv:astro-ph/0506748].
\bibitem{8}A. G. Riess, et al., Type Ia Supernova Discoveries at $z>1$ From the Hubble Space Telescope: Evidence for Past Deceleration and Constraints on Dark Energy Evolution, Astrophys. J. 607(2004)665 [arXiv:astro-ph/0402512].
\bibitem{9}S. Cole, et al., The 2dF Galaxy Redshift Survey: Power-spectrum analysis of the final dataset and cosmological implications, Mon. Not. Roy. Astron. Soc. 362(2005)505 [arXiv:astro-ph/0501174].
\bibitem{10}N. A. Bahcall, J.P. Ostriker, S. Perlmutter, P.J. Steinhardt, The Cosmic Triangle: Revealing the State of the Universe, Science 284(1999)1481.
\bibitem{11}S. M. Carroll, The Cosmological Constant, Living Rev. Rel. 4(2001)1.
\bibitem{12}R. Utiyama and B. S. DeWitt, Renormalization of a Classical Gravitational Field Interacting with Quantized Matter Fields, J. Math. Phys. 3(1962)608.
\bibitem{13}K. S. Stelle, Renormalization of higher-derivative quantum gravity, Phys. Rev. D16(1977)953.
\bibitem{14}N. D. Birrell and P. C. W. Davies, Quantum Fields in Curved Spacetime, Cambridge University Press, Cambridge, 1982.
\bibitem{15}I. L. Buchbinder, S. D. Odintsov, and I. L. Shapiro, Effective Actions in Quantum Gravity, IOP Publishing, Bristol, 1992.
\bibitem{16}G. A. Vilkovisky, Effective action in quantum gravity, Class. Quant. Grav. 9(1992)895.
\bibitem{17}C. Brans and R. H. Dicke, Mach's Principle and a Relativistic Theory of Gravitation, Phys. Rev. 124(1961)925.
\bibitem{18}V. Faraoni, Cosmology in Scalar-Tensor Gravity, Kluwer Academic, Dordrecht, 2004.
\bibitem{19}G. R. Dvali, G. Gabadadze and M. Porrati, 4D Gravity on a Brane in 5D Minkowski Space, Phys. Lett. B485(2000)208.
\bibitem{20}R. Maartens, Brane-World Gravity, Living Rev. Rel. 7(2004)7.

\bibitem{21}J. D. Bekenstein, Relativistic gravitation theory for the MOND paradigm, Phys. Rev. D70(2004)083509.
\bibitem{22}T. Jacobson and D. Mattingly, Gravity with a dynamical preferred frame, Phys. Rev. D64(2001)024028.
\bibitem{23}T. P. Sotiriou and V. Faraoni, f(R) theories of gravity, Rev. Mod. Phys. 82(2010)451 [arXiv:0805.1726v2[gr-qc]].
\bibitem{24}T. V. Ruzmaikina and A. A. Ruzmaikin, Quadratic Corrections to the Lagrangian Density of the Gravitational Field and the Singularity, Zh. Eksp. Teor. Fiz., 57(1969)680, Sov. Phys. JETP 30(1970)372.
\bibitem{25}H. A. Buchdahl, Non-linear Lagrangians and cosmological theory, Mon. Not. R. Astron. Soc. 150(1970)1–8.
\bibitem{26}A. A. Starobinsky, A new type of isotropic cosmological models without singularity, Phys. Lett. B91(1980)99–102.
\bibitem{27}H. J. Schmidt, Fourth order gravity: Equations, history, and applications to cosmology, Int. J. Geom. Meth. Mod. Phys. 4(2007)209–248.
\bibitem{28}A. De Felice, Sh. Tsujikawa, f(R) theories,  Living Rev. Rel. 13(2010)3, [arXiv:1002.4928 [gr-qc]].
\bibitem{29}S. Capozziello, S. Nojiri, S. D. Odintsov, A. Troisi, Cosmological viability of f(R)-gravity as an ideal fluid and its compatibility with a matter dominated phase, Phys. Lett. B639(2006)135-143, [arXiv:astro-ph/0604431].
\bibitem{30}I. L. Buchbinder, S. D. Odintsov and I. L. Shapiro, Effective Actions in Quantum Gravity, IOP Publishing, Bristol, 1992.
\bibitem{31}V. Faraoni, Matter instability in modified gravity, Phys. Rev. D74(2006)104017.
\bibitem{32}R. P. Woodard, Avoiding Dark Energy with 1/R Modifications of Gravity, Lect. Notes Phys. 720(2007)403.
\bibitem{33}P. W. Higgs, Quadratic lagrangians and general relativity, Nuovo Cim. 11(1959)816.
\bibitem{34}B. Whitt, Fourth-order gravity as general relativity plus matter, Phys. Lett. B145(1984)176.
\bibitem{35}G. Magnano, M. Ferraris and M. Francaviglia, Nonlinear gravitational Lagrangians, Gen. Rel. Grav. 19(1987)465.
\bibitem{36}A. Jakubiec and J. Kijowski, On the universality of Einstein equations, Gen. Rel. Grav. 19(1987)719.
\bibitem{37}A. Jakubiec and J. Kijowski, On theories of gravitation with nonlinear Lagrangians, Phys. Rev. D37(1989)1406.
\bibitem{38}A. Jakubiec and J. Kijowski, On theories of gravitation with nonsymmetric connection, J. Math. Phys. 30(1989)1073.
\bibitem{39}G. Magnano, L. M. Sokolowski, On Physical Equivalence between Nonlinear Gravity Theories, Phys. Rev. D50(1994)5039-5059, [arXiv:gr-qc/9312008].
\bibitem{40}J. D. Barrow, S. Cotsakis, Inflation and the conformal structure of higher-order gravity theories, Phys. Lett. B214(1988)515.

\bibitem{41}P. Teyssandier, P. Tourrenc, The Cauchy problem for the $R+R^2$ theories of gravity without torsion, J. Math. Phys. 24(1983)2793.
\bibitem{42}D. Wands, Extended gravity theories and the Einstein-Hilbert action, Class. Quant. Grav. 11(1994)269.

\bibitem{43}E. E. Flanagan, The conformal frame freedom in theories of gravitation, Class. Quant. Grav.21(2004)3817,[arXiv:gr-qc/0403063v3].
\bibitem{44}T. P. Sotiriou, V. Faraoni, S. Liberati, Theory of gravitation theories: a no-progress report, Int. J. Mod. Phys. D17(2008)399-423, [arXiv:0707.2748v2 [gr-qc]].

\bibitem{45}V. Faraoni, Sh. Nadeau, The (pseudo) issue of the conformal frame revisited, Phys. Rev. D75(2007)023501, [arXiv:gr-qc/0612075].
\bibitem{46}R. Catena, M. Pietroni, L. Scarabello, Einstein and Jordan frames reconciled: a frame-invariant approach to scalar-tensor cosmology, Phys.Rev.D76(2007)084039, [arXiv:astro-ph/0604492v2].
\bibitem{47}M. Postma, M. Volponi, Equivalence of the Einstein and Jordan frames, Phys. Rev. D90(2014)103516, [arXiv:1407.6874v2].
\bibitem{48}T. Chiba, M. Yamaguchi, Conformal-Frame (In)dependence of Cosmological Observations in Scalar-Tensor Theory, J. Cosmol. Astropart. Phys. 10(2013)040, [arXiv:1308.1142].
\bibitem{49}I. Quiros, R. Garcia-Salcedo, J. E. M. Aguilar, T. Matos, The conformal transformation's controversy: what are we missing?, Gen. Rel. Grav. 45(2013)489, [arXiv:1108.5857 [gr-qc]].
\bibitem{50}I. Quiros, R. Garcia-Salcedo, J. E. M. Aguilar, Conformal transformations and the conformal equivalence principle, arXiv:1108.2911[gr-qc].
\bibitem{51}C. Romero, J. B. Fonseca-Neto, M. L. Pucheu, General Relativity and Weyl Frames, arXiv:1106.5543 [gr-qc].
\bibitem{52}F. Bezrukov, M. Shaposhnikov, Standard Model Higgs boson mass from inflation: two loop analysis, JHEP, 0907 (2009)089,[arXiv:0904.1537 [hep-ph]].
\bibitem{53}A. De Simone, M. P. Hertzberg, F. Wilczek, Running Inflation in the Standard Model, Phys. Lett. B678(2009)1, [arXiv:0812.4946 [hep-ph]].
\bibitem{54}A. O. Barvinsky, A. Y. Kamenshchik, A. A. Starobinsky, Inflation scenario via the Standard Model Higgs boson and LHC, JCAP 0811(2008)021, [arXiv:0809.2104 [hep-ph]].
\bibitem{55}F. Briscese, E. Elizalde, S. Nojiri, S. D. Odintsov, Phantom scalar dark energy as modified gravity: Understanding the origin of the Big Rip singularity, Phys. Lett. B646(2007)105, [arXiv:hep-th/0612220].
\bibitem{56}J. White, M. Minamitsuji, M. Sasaki, Curvature perturbation in multi-field inflation with non-minimal coupling, JCAP 1207(2012)039, [arXiv:1205.0656[astro-ph.CO]].
\bibitem{57}C. H. Brans, Nonlinear Lagrangians and the significance of the metric, Class. Quant. Grav. 5(1988)L197.
\bibitem{58}V. Faraoni, E. Gunzig, P. Nardone, Conformal transformations in classical gravitational theories and in cosmology, Fund. Cosmic Phys. 20(1999)121, [arXiv:gr-qc/9811047v1].
\bibitem{59}S. Capozziello, P. Martin-Moruno, C. Rubano, Physical non-equivalence of the Jordan and Einstein frames, Phys. Lett. B689(2010)117–121, [arXiv:1003.5394].
\bibitem{60}V. Faraoni, E. Gunzig, Einstein frame or Jordan frame , Int. J. Theor. Phys. 38(1999)217-225, [arXiv:astro-ph/9910176].
\bibitem{61}C. H. Brans, The roots of scalar-tensor theory: an approximate history, arXiv:gr-qc/0506063.
\bibitem{62}C. Schlippert, et al., Quantum Test of the Universality of Free Fall, Phys. Rev. Lett. 112(2014)203002, [arXiv:1406.4979 [physics.atom-ph]].
\bibitem{63}Ta-Pei Cheng, Relativity, Gravitation and Cosmology, second ed., Oxford University Press Inc., New York, 2010, pp.199-200.
\bibitem{64}S. Bahamonde, S. D. Odintsov, V. K. Oikonomou, M. Wright, Correspondence of F(R) gravity singularities in Jordan and Einstein frames, Annals of Phys. 373(2016)96-114.
\bibitem{65}R. H. Dicke, Mach's Principle and Invariance under Transformation of Units, Phys. Rev. 125(1962)2163.
\bibitem{66}C. Wetterich, A Universe without expansion, Physics of the dark universe, 2(2013)184, [arXiv:1303.6878].
\bibitem{67}E. Poisson, The Motion of Point Particles in Curved Spacetime, Living Rev. Relativity, 7(2004)6.
\bibitem{68}Sh. Nojiri, S. D. Odintsov, Unified cosmic history in modified gravity: from F(R) theory to Lorentz
non-invariant models, Phys.Rept. 505(2011)59-144, [arXiv:1011.0544].
\bibitem{69}Sh. Nojiri, S. D. Odintsov, Introduction to modified gravity and gravitational alternative for dark
energy, Int. J. Geom. Meth. Mod.Phys. 4(2007)115-146, [arXiv:hep-th/0601213].
\bibitem{70}Sh. Nojiri, S. D. Odintsov,  V. K. Oikonomou, Modified Gravity Theories on a Nutshell: Inflation, Bounce and Late-time
Evolution, Phys.Rept. 692(2017)1-104, [arXiv:1705.11098].
\bibitem{71}S. Bahamonde, S. D. Odintsov, V. K. Oikonomou, P. V. Tretyakov, Deceleration versus acceleration universe in different frames of F(R) gravity, Phys.Lett. B766(2017)225-230, [arXiv:1701.02381].
\bibitem{72}D. J. Brooker, S. D. Odintsov, R. P. Woodard, Precision predictions for the primordial power spectra from  f(R) models of inflation, Nucl. Phys. B911(2016)318-337, [arXiv:1606.05879].
\bibitem{73}Sk. Nayem, A. K. Sanyal, Why scalar-tensor equivalent theories are not physically equivalent?, Int. J. Mod. Phys. D26(2017)1750162, [arXiv:1609.01824 [gr-qc]].
\bibitem{74}S. Chakraborty, S. SenGupta, Solving higher curvature gravity theories, Eur. Phys. J. C76 (2016) no.10, 552, [arXiv:1604.05301].
\end{thebibliography}
\end{document}